\documentclass{amsart}
\usepackage[foot]{amsaddr}
\usepackage{hyperref}
\usepackage[super,sort,compress]{cite}
\usepackage[ruled,linesnumbered,lined]{algorithm2e}
\usepackage[braket]{qcircuit}
\usepackage{color,colortbl,multirow}
\usepackage{pgfplots}
\usepackage{tikz,pgfplots}
\usepackage{pgfplotstable}
\usepackage[caption=false]{subfig}
\usepackage{amsmath, amssymb}
\usepackage{booktabs}

\pgfplotsset{width=10cm,compat=1.9}

\begin{document}

\title[]{Quantum enhanced cross-validation for near-optimal neural networks architecture selection}
\author{Priscila G. M. dos Santos, Rodrigo S. Sousa, Ismael C. S. Araujo and Adenilton J. da Silva}

\address{
Departamento de Computa\c{c}\~{a}o\\
Universidade Federal Rural de Pernambuco\\
Rua Dom Manoel de Medeiros, s/n. Campus Dois Irm\~aos\\
52171-900, Recife, Pernambuco, Brazil}
\email{\{priscila.marques, rodrigo.silvasouza, ismael.cesar, adenilton.silva\}@ufrpe.br}

\begin{abstract}
This paper proposes a quantum-classical algorithm to evaluate and select classical artificial neural networks architectures.
The proposed algorithm is based on a probabilistic quantum memory and the possibility to train artificial neural networks in superposition.
We obtain an exponential quantum speedup in the evaluation of neural networks.
We also verify experimentally through a reduced experimental analysis that the proposed algorithm can be used to select near-optimal neural networks.
\end{abstract}

\maketitle

\section{Introduction}
\label{sec:intro}

Artificial neural networks (ANN) are computational models inspired by the human brain and with learning capacities.
The first artificial neuron was proposed in the 1940s\cite{mcculloch1943logical}, a learning rule is proposed by Hebb~\cite{hebb2005organization} and
the backpropagation algorithm based on gradient descent was proposed in 1980s~\cite{rumelhart1988learning}.
ANNs have several applications in industry and research.
For instance, in pattern recognition\cite{samarasinghe2016neural}, clustering\cite{xu2017self}, image\cite{gatys2016image} and speech processing\cite{chan2016listen} and other applications.

An artificial neuron with $m$ real inputs $x_1, \dots, x_m$ has $m$ weights $w_1, \dots, w_m$, a bias $b$ and its output is described in Eq.~\eqref{eq:neuron}, where $f$ is a nonlinear activation function.
\begin{equation}\label{eq:neuron}
  f\left(\sum_{k=1}^m w_k \cdot x_k + b\right)
\end{equation}

A Feedforward Neural Network (FNN) is composed of layers of neurons and each layer receives its input signal from the previous layer.
FNN optimization has received much attention in the last 20 years~\cite{ojha2017metaheuristic}.
Metaheuristics as meta-learning\cite{abraham2004meta}, differential evolution\cite{ilonen2003differential}, genetic algorithms\cite{montana1989training},
evolutionary programming, simulated annealing, tabu search\cite{pham2012intelligent}, particle swarm optimization\cite{zhang2007hybrid}, etc.\cite{ojha2017metaheuristic} have been
used to optimize neural networks architecture.

The number of neurons in the hidden layers and the number of hidden layers are some of the most important elements of a feedforward ANN because there is a relation between them and the ANN
performance\cite{benardos2007optimizing}.
The optimization of neural networks weights with backpropagation or
other techniques based on gradient descend leads to local minima in the error space.
To evaluate a neural network architecture, it is necessary to perform an empirical evaluation that involves a tedious
trial and error process with several random weights initializations.
This trial and error process can involve a procedure to estimate the accuracy of candidate classifiers.
The $\kappa$-fold cross-validation\cite{kohavi1995} is an accuracy estimation method used, for instance, to perform model evaluation and model selection.
A dataset $T$ is split in $\kappa$ disjoint folds or subsets $T_1, \dots, T_\kappa$ and
a classifier is trained $\kappa$ times in which each iteration $t \in [1,\kappa]$ the model is trained with dataset $T-T_t$ and tested with fold $T_t$.

To determine if a neural network architecture can learn a given task is an NP-complete problem named the loading problem~\cite{judd1990neural}.
If $P \neq NP$ then developing a function that maps neural networks architectures to their best performance over a given data set
is an intractable problem. The objective of this work is to investigate the possibility to use quantum computation
for selecting a near-optimal classical neural network architecture for a given learning task.
In previous works on neural network architecture evaluation~\cite{panella2011neural} or architecture selection on a quantum computer\cite{da2016quantum},
a nonlinear quantum operator was used to propose a polynomial algorithm that solves the loading problem.
As it is not known whether  nonlinear quantum operators are physically realizable or not,
in this paper we take the safer road by obeying the principles of quantum mechanics by using unitary quantum operators.
We have already followed this track by performing
an \textit{evaluation} of neural networks performances using unitary quantum operators\cite{silva2017quantum}, here we address the problem of unitarily performing an \textit{architecture selection} of neural networks.

Several quantum machine learning models\cite{biamonte2017quantum} and  quantum neural networks~\cite{schuld2014quest} have been proposed,
but the non-existence\footnote{Actual quantum computers do not have enough quantum bits, ``remain coherent for a limited time\cite{cross2017open}'' or are designed for specific tasks.
The quantum computer necessary to perform the tasks described in this work should be universal and have thousands of qubits  } of quantum computers does not allow an empirical comparison between classical and
quantum neural networks models.
We cannot evaluate numerically the quantum proposed models to verify if they present advantages when compared with
classical models. This technical limitation is named the benchmark problem~\cite{biamonte2017quantum}.
The algorithm proposed in this work is a quantum algorithm and requires a universal quantum computer. It is also designed to allow a (reduced) simulation in a classical computer and
we show that the proposed method can choose a near optimal neural network architecture without
the necessity of random weights initializations and with a single training of each neural network architecture.
This result has two main implications: i) we can use a quantum enhanced cross-validation to perform neural network parameter evaluation/selection with an exponential quantum speedup
and ii) the proposed method can be evaluated numerically and presents advantages over classical strategies using real benchmark problems.

The remainder of this work is organized into 5 sections.
Section~\ref{sec:qam} presents the probabilistic quantum memory used in this work.
Section~\ref{sec:nnselection} is the main section and presents a quantum algorithm that evaluates classical neural networks architectures and
is used to perform neural networks architecture selection. Section~\ref{sec:exp} presents experiments,  that have been performed in a classical computer (and can be executed exponentially faster in a quantum computer). Section~\ref{sec:discussion} presents a discussion of the results.
Section~\ref{sec:conclusion} presents the conclusion.

\section{Probabilistic quantum memories}
\label{sec:qam}
	A content-addressable memory is called associative memory because of the possibility to retrieve information from it even with partial knowledge of the desired content.
	Models of associative memories, like the Hopfield network,
	suffer from a capacity shortage~\cite{trugenberger2001probabilistic}.
	The quantum counterpart of an associative memory has the advantage of having an exponential capacity because the patterns stored in the memory are kept in superposition. Given a dataset of $n$ patterns with $k$ qubits $T=\{p^1, p^2,...,p^n\}$, the quantum memory creates the state described in Eq. \eqref{eq:quamtumMemory},
	where $\ket{M}$ is the quantum register that will store the patterns.
	\begin{equation}
		\ket{M} = \frac{1}{\sqrt{n}}\sum_{j = 1}^n\ket{p^j}
		\label{eq:quamtumMemory}
	\end{equation}

In this work, we use the Probabilistic Quantum Memory\cite{trugenberger2002quantum} (PQM). 
The storage algorithm of the PQM creates a superposition of binary patterns as described in Eq.~\eqref{eq:quamtumMemory}. The retrieval algorithm of the PQM is probabilistic and depends on the Hamming distance between the input pattern and stored patterns.

It is necessary to reload the memory after each execution of the recovering algorithm of the probabilistic quantum memory. 
This problem is pointed out as a fundamental limitation of the PQM\cite{PhysRevLett.91.209801} and decreases the speedup for tasks as machine learning\cite{schuld2014quantum}. 
In this work, we propose an application of the PQM that requires a single execution of the PQM recovering algorithm for a given input and this limitation does not affect the method proposed in this paper.

One second limitation of the PQM is its inability to deal with continuous inputs. In the actual small-scale quantum computers this is a strong limitation, but if quantum computers with thousands or millions of qubits are built the binary representation can be used to represent continuous inputs with some precision. In this work, we assume the existence of such quantum computers and continuous inputs can be approximately represented by using binary numbers.

\subsection{The storage algorithm}
	The states during the PQM storage algorithm are divided into three quantum registers $\ket{p_1p_2...p_k;u_1u_2;m_1m_2...m_k}$.
Where $\ket{p_j}$ is the $j$-th qubit of the input register,
$u_1u_2$ are control qubits prepared in a state $\ket{01}$ and $\ket{\textbf{m}} =\ket{m_1,\dots, m_k}$ is the memory register, where the patterns are to be stored.
	To build a coherent superposition of the patterns to be stored it is necessary to make use of $Toffoli$, $X$ and $CS^j$ gates. The $CS^j$ gate is described in Eq.~\ref{eq:csGate}.

\begin{equation}
CS^{j} =
	\left[\begin{array}{cccc}
		1 & 0 &      0                 &          0\\
		0 & 1 &      0                 &          0 \\
		0 & 0 & \sqrt{\frac{j-1}{j}}   &  \frac{1}{\sqrt{j}} \\
		0 & 0 & \frac{-1}{\sqrt{j}}    &  \sqrt{\frac{j-1}{j}}
	\end{array}\right]
	\label{eq:csGate}
\end{equation}

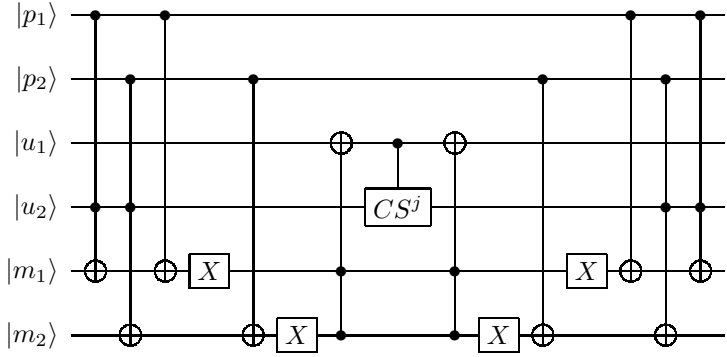
\begin{figure}
	$$
\Qcircuit @C=.5em @R=1.0em @!R{
\lstick{\ket{p_1}}&\ctrl{3}&\qw     &\ctrl{4}&\qw     &\qw     &\qw     &\qw      &\qw        &\qw      &\qw     &\qw     &\qw     &\ctrl{4}&\qw     &\ctrl{3}&\qw\\
\lstick{\ket{p_2}}&\qw     &\ctrl{2}&\qw     &\qw     &\ctrl{4}&\qw     &\qw      &\qw        &\qw      &\qw     &\ctrl{4}&\qw     &\qw     &\ctrl{2}&\qw     &\qw\\
\lstick{\ket{u_1}}&\qw	   &\qw     &\qw     &\qw     &\qw     &\qw     &\targ    &\ctrl{1}   &\targ    &\qw     &\qw     &\qw     &\qw     &\qw     &\qw     &\qw\\
\lstick{\ket{u_2}}&\ctrl{1}&\ctrl{2}&\qw     &\qw     &\qw     &\qw     &\qw      &\gate{CS^j}&\qw      &\qw     &\qw     &\qw     &\qw     &\ctrl{2}&\ctrl{1}&\qw\\
\lstick{\ket{m_1}}&\targ   &\qw     &\targ   &\gate{X}&\qw     &\qw     &\ctrl{-2}&\qw        &\ctrl{-2}&\qw     &\qw     &\gate{X}&\targ   &\qw     &\targ   &\qw\\
\lstick{\ket{m_2}}&\qw     &\targ   &\qw     & \qw    &\targ   &\gate{X}&\ctrl{-1}&\qw        &\ctrl{-1}&\gate{X}&\targ   &\qw     &\qw     &\targ   &\qw     &\qw
}
	$$
	\caption{ A quantum circuit for storing patterns of two qubits}
	\label{fig:store}
\end{figure}

A circuit representing a 2 qubit probabilistic quantum memory storing procedure is described in Fig.~\ref{fig:store}. 
In the first iteration the quantum registers $\ket{p,u,m}$ are initialized as described in the Eq.~\ref{eq:psi0} and run the circuit described in Fig.~\ref{fig:store}. 
For each other pattern $p^j$ in the dataset, we initialize the quantum register input with $p^j$ and execute the circuit described in Fig. \ref{fig:store} again.

\begin{equation}
	\ket{\psi_0} = \ket{p_1^1p_2^1...p_k^1,01,00...0}.
	\label{eq:psi0}
\end{equation}

In order to store the $n$ patterns of the set $P$, it is necessary  to perform $n$ iterations.
In the application of the $CS^j$ gate, the parameter $j$ is $j=p+1-iter$ where $p$ is the number of patterns and
$iter$ is the iteration number.

\subsection{The retrieval algorithm}
The retrieval algorithm also requires three registers~\cite{trugenberger2001probabilistic,trugenberger2002quantum} $\ket{i_1...i_k;m_1...m_k;c_1...c_d}$.
Where $\ket{i_j}$ is the $j$-th qubit of the input register, $\ket{m_j}$ is the $j$-th qubit of the memory register and $c_j$ is the $j$-th qubit of the control register~\cite{trugenberger2002quantum},
The control qubits start at the state $\ket{0...0}$.
The retrieval of the patterns is made probabilisticaly~\cite{trugenberger2001probabilistic}.
For each $c_l$ in $\ket{c}$, the following steps are performed. Step 1 apply the Hadamard gate to $\ket{c_l}$ giving us, at the first iteration, the state described in Eq.\eqref{eq:step1pqm}.

\begin{equation}
	\ket{\psi_0} = \frac{1}{\sqrt{2n}} \sum_{j=1}^n \ket{i_1...i_k;p_1^j...p_k^j;0_10_2...0_b} + \frac{1}{\sqrt{2n}} \sum_{j=1}^n\ket{i_1...i_k;p_1^j...p_k^j;1_10_2...0_d}
	\label{eq:step1pqm}
\end{equation}

In the second step, for each bit $i_j$ in the input pattern, 
we apply the CNOT gate with the $j$-th qubit of the input as the control
and the $j$-th qubit of the memory as the target and apply the $NOT$ gate to the $j$-th bit of the memory to obtain the state 
$$\ket{\psi_1} = \prod_{j=1}^n CNOT(i_j, m_j) X(m_j) \ket{\psi_0}.$$
After the second step,
if there is a pattern stored in the superposition equal to the input all its qubits will be set to $\ket{1}$~\cite{trugenberger2001probabilistic}.

Step 3 applies the quantum operator described in Eq.~\ref{eq:pqm4}.
\begin{equation}
	\ket{\psi_2} = \prod_{j=1}^k(CV^{-2})(c_l,m_j)\prod_{j=1}^k U(m_j)\ket{\psi_1}
	\label{eq:pqm4}
\end{equation}
Where the operator $V$ is a unitary matrix and $CV^{-2}$ is a controlled version of the $V^{-2}$ operator. 
$$
	V = \left[\begin{array}{cc}
	     e^{(i\frac{\pi}{2n})} &   0 \\
			  0                  &   1
			\end{array}\right]
$$

The inverse of steps 1 and 2 are applied to the quantum state $\ket{\psi_2}$ to restore the memory quantum register to its original state. This is the last deterministic step of the retrieval algorithm and the resulting state is described in Eq.~\eqref{eq:pqm}.

\begin{equation}
	\ket{\psi} = \frac{1}{\sqrt{n}}\sum_{j=1}^n\sum_{l=0}^d cos^{d-l}\left(\frac{\pi}{2k}d_h\left(i,p^j\right)\right)\cdot
sen^{l}\left(\frac{\pi}{2k}d_h\left(i,p^j\right)\right)\sum_{\left\{J^l\right\}}\ket{i;p^k;J^l}
\label{eq:pqm}
\end{equation}

Where $\left\{J^l\right\}$ is the set of all binary strings with exactly $l$ bits set to $1$ and $\left(d-l\right)$ bits set to $0$~\cite{trugenberger2002quantum}.
After processing the state, it is necessary to perform a measurement to the control qubits.

The result of the measurement is a large number of control qubits in the state $\ket{0}$
if all stored patterns are similar to the input, and a large number of control qubits in the state $\ket{1}$ if all stored patterns are very distant of the input. In this work, we consider that the number of $1$s obtained after the measurement of quantum register $\ket{c}$ is the output $y$ of the PQM.
From Eq.~\ref{eq:pqm} we can easily verify that the probability to obtain $y = \mathcal{K}$ is given by
\begin{equation}
P(y = \mathcal{K}) = \frac{1}{p}\sum_{j=1}^p\binom{d}{\mathcal{K}} cos^{2\left(d-\mathcal{K}\right)}\left(\frac{\pi}{2k}d_h\left(i,p^j\right)\right)
\cdot
sin^{2\mathcal{K}}\left(\frac{\pi}{2k}d_h\left(i,p^j\right)\right)
\end{equation}

\section{Selection of neural networks architecture in a quantum computer}
\label{sec:nnselection}
An artificial neural network for classification is defined as a function $N: \mathbb{R}^m \rightarrow \{c_1, \dots, c_k\}$.
Despite being a real function the implementation of a neural network in a classical digital computer is a binary function.
All binary functions can be simulated in a quantum computer, then theoretically a classical neural network can be represented by a quantum circuit where
the weights, inputs and outputs are strings of qubits.
One backpropagation iteration is also a real function that receives inputs $x(t)$ and weights $w(t)$, and outputs weights $w(t+1)$.
In a digital computer one backpropagation iteration is also a binary function and theoretically can be represented with a quantum operator.

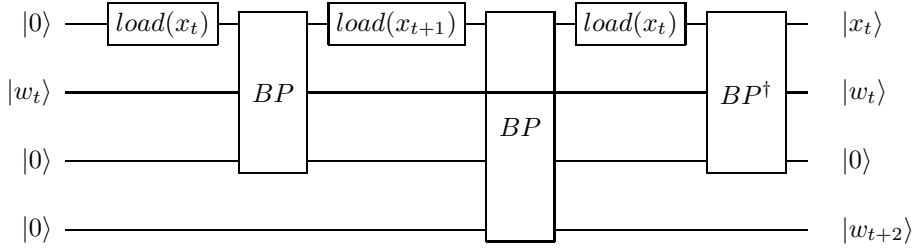
\begin{figure}
  \[
  \Qcircuit @C=0.8em @R=1.0em @!R{
  \lstick{\ket{0}}     	& \qw & \gate{load(x_t)} & \multigate{2}{BP} & \gate{load(x_{t+1})} & \multigate{3}{BP} & \gate{load(x_t)} & \multigate{2}{BP^{\dag}} & \qw & \rstick{\ket{x_t}} \\
  \lstick{\ket{w_t}} 	& \qw & \qw & \ghost{BP}        & \qw & \qw               & \qw & \ghost{BP^{\dag}}            &\qw & \rstick{\ket{w_t}} \\
  \lstick{\ket{0}}   	& \qw & \qw & \ghost{BP}        & \qw & \ghost{BP}        & \qw & \ghost{BP^{\dag}}          & \qw & \rstick{\ket{0}}   \\
  \lstick{\ket{0}}   	& \qw & \qw & \qw               & \qw & \ghost{BP}        & \qw & \qw                        & \qw & \rstick{\ket{w_{t+2}}}
  }
  \]
  \caption{Theoretical quantum circuit implementing two backpropagation iterations }
  \label{fig:qbackp}
\end{figure}

Figure~\ref{fig:qbackp} represents the idea of training neural networks in superposition. The first quantum register receives patterns from the training set. The load function loads a pattern in the quantum register. This loading function can be accomplished because the state of input quantum register is always a basis state.
The BP operator represents a backpropagation step.
The first BP operator receives input $\ket{x_t}$ and weights $\ket{w_t}$ and produces $\ket{w_{t+1}}$.
The second BP operator receives $\ket{x_{t+1}}$ and $\ket{w_{t+1}}$ to produce $\ket{w_{t+2}}$.
The quantum operator $BP^{\dag}$ inverts the action of the first $BP$ operator and
prepares the third quantum register to receive the next weight vector.
If $\ket{w_t}$ is a superposition of weights,
a sequence of $load$, $BP$ and $BP^\dag$ operators will train all the networks in the superposition simultaneously.
We can obtain all the neural networks with a given architecture in superposition
just by applying the Hadamard operator in all qubits in the quantum register $\ket{w}$.

The idea to present an input pattern to neural networks in superposition is presented by Meener~\cite{menneer1995quantum}, he named this strategy Strongly Inspired Neural Network.
Instead of developing the strongly inspired neural network the authors developed a weakly inspired neural network,
where for each pattern in a dataset a neural network is trained and later all neural networks are stored in superposition.
We develop the strategy of strong inspired neural networks and we use the name superposition based learning\cite{ricks2004training} to avoid confusion with works that deal with
classical neural networks only based on ideas from quantum computing.

In addition to creating a superposition of neural networks, we also manage the dataset to perform a $\kappa$-fold cross-validation with the neural networks in superposition.
To perform the cross-validation using quantum superposition,
we use two additional quantum registers $\ket{fold}$ representing the fold used as test and $\ket{input\_fold}$ which contains information about the fold of the actual input sample.
The learning algorithm iteration is applied only when $\ket{fold}$ and $\ket{input\_fold}$ are different.
If the $\ket{fold}$ quantum register receives a superposition representing all possible folds, the cross-validation can be performed in superposition with the cost of only one learning algorithm execution.

The effect to train all neural networks in superposition is to obtain a superposition of weight vectors at local minima of the error surface.
Classically one could just choose the neural network with the best accuracy over a validation dataset.
But this information cannot be accessed directly from the state in quantum superposition.
To obtain a useful measure of the architecture performance for the dataset
we use a quantum procedure to calculate the distance between the neural networks accuracy and 100\% of accuracy in the validation set.

We calculate the performance of the neural networks in superposition by presenting patterns in fold $l$ $(l=1, \dots, \kappa)$  to the neural networks in superposition
and applying an $X$ gate in the $ith$ qubit of the performance quantum register if the network correctly classifies the $ith$ pattern and $testFold$ is equal to $l$.
After this procedure the performance and weights quantum registers will be entangled and their state is described in Eq.~\eqref{eq:perf}.
\begin{equation}
  \sum_{w, testFold} \ket{testFold}\ket{w}\ket{performance_{w,testFold}}
  \label{eq:perf}
\end{equation}
It is necessary to present the dataset only once to calculate the performance of all neural networks in all folds.

The quantum register performance is in state $$\sum_{w, testFold} \ket{performance_{w, testFold}},$$
we use this state as the memory of a probabilistic quantum memory with
input $\ket{1}_n$ with $k$ control qubits. The auxiliary quantum register $\ket{c}$ is measured and
the output is the number of 1s obtained. The algorithm is repeated one time for each architecture to be evaluated.

Algorithm~\ref{alg:nnarchs} presents the quantum algorithm to select neural networks architectures. We suppose that the quantum device is controlled by a host classical computer. Heterogeneous computer architectures with different processors specialized for different tasks are increasingly common. The first quantum computers with such architecture are in development. Algorithm~\ref{alg:nnarchs} uses this hybrid architecture to perform a heuristic search over classical neural networks architectures using a quantum device.

Step 1 of Algorithm~\ref{alg:nnarchs} creates the folds of the cross-validation and is performed in the classical computer. Each fold created in the cross-validation procedure has the same size and if necessary some patterns are removed from the dataset.

The for loop starting in step 2 performs an enhanced cross-validation of all neural networks with a given architecture. Steps 3 to 5 are initialization steps and are executed in the theoretical quantum device. Step 3 initializes neural network weights in a superposition of all possible weights. Step 4 initializes the fold quantum register with the quantum state $\sum_{l=1}^\kappa \frac{1}{\sqrt{\kappa}}\ket{l}$. Step 5 initializes the quantum register performance with the quantum state $\ket{0}_n$. After step 5, the state of quantum registers testFold, weights and performance will be described by Eq.~\eqref{eq:s5}, where $W$ is the set of all possible weights with a given precision.

\begin{equation}
    \ket{testFold}\ket{weights}\ket{performance} =
    \sum_{\substack{w \in W, \\ testFold \in \{1, \cdots, \kappa\}}}
    \ket{testFold}\ket{w}\ket{0}_n
    \label{eq:s5}
\end{equation}

Step 6 trains the neural networks in superposition.
The training procedure is a quantum-classical algorithm described in Fig.~\ref{fig:qbackp}.
At each iteration, the classical device selects a pattern from the dataset and loads the pattern in the input quantum register (that is always in a basis state) and the fold of the pattern in the inputFold quantum register.
Then the learning iteration is performed in the parcels in the superposition where the testFold and inputFold are different. After this step the weights quantum register is in a superposition of trained neural networks and the testFold, weights and performance quantum register will be in the state described in Eq.~\eqref{eq:trained} where $\tilde{W}_{testFold}$ is the set of neural networks weights trained with data $T-T_{testFold}$.

\begin{equation}
    \sum_{\substack{w_{testFold} \in \tilde{W}_{testFold}, \\ testFold \in \{1, \cdots, \kappa\}}}
    \ket{testFold}\ket{w_{testFold}}\ket{0}
    \label{eq:trained}
\end{equation}

The for loop starting in step 7 calculates the performance of each neural network in superposition in the validation set. Each test fold of the cross-validation is presented and the neural networks accuracy correspondent to this fold is evaluated. The evaluation is performed in superposition and the test set needs to be presented only once. After this for loop, the state of quantum registers testFold, weights and performance are described in Eq.~\eqref{eq:evaluate}.

\begin{equation}
    \sum_{\substack{w_{testFold} \in \tilde{W}_{testFold}, \\ testFold \in \{1, \cdots, \kappa\}}}
    \ket{testFold}\ket{w_{testFold}}\ket{performance_{w_{testFold}}}
    \label{eq:evaluate}
\end{equation}

Step 15 runs the recovering algorithm of the probabilistic quantum memory with input $\ket{1}_n$ representing a performance of 100\% and the state in quantum register performance as memory.

Step 16 measure the output of the quantum probabilistic memory and store the number of 1s in a classical variable $n_N$ for each architecture $N$. At the end of the algorithm the simplest network $N$ that minimizes the value of $n_N$ is indicated by the algorithm.

\begin{algorithm}
  Divide the dataset $\mathcal{T}$ in $\kappa$ folds \label{line:2}\\
\For{\textbf{each} neural network architecture $N$}
{
  Initialize all weights qubits with $H\ket{0}$  \label{line:1}\\
  Create a superposition with the values 1 to $\kappa$ in quantum register testFold\\
  Initialize quantum register $\ket{performance}$ with the quantum register $\ket{0}_n$ \label{line:3}\\
  Train the neural networks in superposition with the folds with label different of testFold \label{line:4}\\
    \For{\textbf{each} pattern $p_j$ and desired output $d_j$ in $testFold_j$ \label{line:5}}{
    Initialize the quantum registers input, calculatedOutput and desiredOutput with the basis quantum state $\ket{p_j,0,d_j}$ \label{line:6}\\
    Calculate $N\ket{p_k}$ to calculate network output in quantum register calculatedOutput \ket{o} \label{line:7}\\
      \If{$\ket{o} = \ket{d}$ and $\ket{testFold} = \ket{inputFold(p_j)}$  \label{line:8}}
  	   {Set $\ket{performance}_j$ to 1 \label{line:9}}
    Calculate $N^{-1}$ to restore $\ket{o}$ \label{line:10}
    }\label{line:11}
  Apply the quantum associative recovering algorithm with input $\ket{1}_n$, memory $\ket{performance}$ and
  $b$ qubits in the output  \label{line:13}\\
  Measure quantum register $\ket{c}$ and stores the number of 1s in  $n_N$
}
  Return the simplest neural network architecture $N$ that minimize $n_N$.
\caption{Architecture selection}
\label{alg:nnarchs}
\end{algorithm}

\section{Experiments}
\label{sec:exp}

Since there are no quantum computers with sufficient qubits to run
the proposed algorithm, it was necessary to perform some changes in
the quantum algorithm in order to simulate it on a classical computer.
Therefore, we reduced (without loss of generality) the number of neural network instances in the
quantum parallelism. Besides this change, we simply followed the algorithm description
in order to make a classical version of Algorithm~1.

To perform the experiments, we use the Multilayer
Perceptron (MLP). The training and
evaluation were performed using the scikit-learn\cite{pedregosa2011scikit}
version 0.19.1. After training the neural networks we evaluated the performance of every neural
network instance and stored the performance vectors which have the size of the validation set and the $i$-th position is set to 1 if the trained network correctly classifies the $i$-th vector in the validation set and is set to 0 otherwise.
The performance vectors are used as the memory of the probabilistic quantum memory and the output of the probabilistic quantum memory was calculated using Eq.~\eqref{eq:pqm}.

The datasets used in this work were from the PROBEN1 repository, which consists in a collection
of problems for neural network learning in the realm of pattern classification
and function approximation\cite{prechelt1994proben1}. PROBEN1 contains 15 datasets
from real-world problems and from 12 different domains. We used 8 datasets to perform
the experiments: cancer, gene, diabetes, card, glass, heart, horse
and mushroom. The details about the datasets used can be seen in Table~\ref{tab1}.
The datasets were divided into 10 folds and  the train set contains 9 folds
 while the test set contains the remaining fold. The number of
output qubits in the probabilistic quantum memory was set to 100.

\begin{table}[t]
\caption{Datasets.\label{tab1}}
{\begin{tabular}{@{}ccccc@{}} \toprule
Dataset & features & classes & examples & description \\ \hline
  cancer & 9 & 2 & 699 & diagnosis of breast cancer \\
  gene & 120 & 3 & 3175 & detect intron/exon boundaries in nucleotide sequences \\
  diabetes & 8 & 2 & 768 & diagnose diabetes of Pima indians \\
  card & 51 & 2 & 690 & predict the approval of a credit card to a customer \\
  glass & 9 & 6 & 214 & classify glass types \\
  heart & 35 & 2 & 920 & predict heart disease \\
  horse & 58 & 3 & 364  & predict the fate of a horse that has a colic \\
  mushroom & 125 & 2 & 8124 & discriminate edible from poisonous mushrooms \\ \hline
\end{tabular}}
\end{table}

We consider the number of neurons in the hidden layer as
the architecture to be evaluated. Thus, the number of neurons in the hidden
layer was varied between 1 and 20. All neural network architectures were
trained and tested for every dataset and for 1000 different  initialization
weights (100 for each fold). The alpha parameter used avoids overfitting by constraining the size
of the weights. The learning algorithm is `adam' which refers to the stochastic
gradient-based optimizer proposed by Kingma, Diederik,
and Jimmy Ba\cite{kingma2014adam}. The parameters used can be seen in Table~\ref{tab2}.

\begin{table}[t]
\caption{MLP Parameters.\label{tab2}}
{\begin{tabular}{@{}cccc@{}} \toprule
Parameter & Value & Description \\  \hline
  solver & adam & stochastic gradient-based optimizer \\
  alpha & 1e-4 & regularization term parameter \\
  beta\_1 & 0.9 & decay rate for estimates of first moment vector\\
  beta\_2 & 0.999 & decay rate for estimates of second moment vector\\
  epsilon & 1e-8 & value for numerical stability in adam\\
  max\_iter & 100 & maximum number of iterations \\
  activation & relu & rectified linear unit function \\
  learning\_rate\_init & 1e-3 & controls the step-size in updating the weights \\
  number of hidden neurons & [1,20] & number of hidden neurons  \\ \hline
\end{tabular}}
\end{table}

\section{Results and discussion}
\label{sec:discussion}

\pgfplotstableread{
X Y label
0.526857142857 46.3695362487 a
0.714828571429 25.9857573853 a
0.842314285714 12.7965761169 a
0.893028571429 7.82766845602 a
0.933614285714 3.99028462548 a
0.956585714286 1.83397985345 a
0.9647 1.12132125907 a
0.973014285714 0.420021963266 a
0.9733 0.494513482368 a
0.975342857143 0.305882269672 a
0.976328571429 0.15813923668 a
0.976357142857 0.157136352351 a
0.976571428571 0.154271152978 a
0.9773 0.146073058946 a
0.976542857143 0.154270646429 a
0.976885714286 0.15104987925 a
0.976971428571 0.148839287813 a
0.977428571429 0.142605219962 a
0.977457142857 0.142303646678 a
0.977242857143 0.144868590039 a
0.437043478261 56.6120082649 b
0.573217391304 39.126139294 b
0.655289855072 27.9059631687 b
0.713260869565 20.0207706586 b
0.739188405797 16.6325519953 b
0.757463768116 14.2591497754 b
0.769826086957 12.8172844783 b
0.778202898551 11.8498453914 b
0.781449275362 11.4873780767 b
0.78515942029 11.1143781022 b
0.78715942029 10.905145981 b
0.791376811594 10.4839113128 b
0.792304347826 10.3934961585 b
0.796405797101 10.0005140361 b
0.797289855072 9.92539391697 b
0.799217391304 9.73407181314 b
0.800217391304 9.65292965102 b
0.80168115942 9.51309941514 b
0.80431884058 9.27563133842 b
0.805173913043 9.19331216074 b
0.392220779221 60.5364422302 c
0.475285714286 51.5940789981 c
0.533688311688 44.1669819148 c
0.589467532468 36.5761393108 c
0.633779220779 30.3827642179 c
0.668714285714 25.5186969573 c
0.692584415584 22.3020629007 c
0.717454545455 19.0426329371 c
0.732701298701 17.098702409 c
0.743220779221 15.8308513359 c
0.755311688312 14.3808336083 c
0.762818181818 13.5260594596 c
0.768064935065 12.9190465491 c
0.773025974026 12.3630489169 c
0.776649350649 11.9612065936 c
0.777272727273 11.8812599959 c
0.781480519481 11.4429699345 c
0.781493506494 11.4315760313 c
0.782350649351 11.3380292408 c
0.782467532468 11.322901234 c
0.57777672956 38.3818856718 d
0.786220125786 13.2412398149 d
0.861858490566 5.11634618545 d
0.878059748428 3.73193654829 d
0.880946540881 3.51523466232 d
0.882933962264 3.38767793498 d
0.882650943396 3.39400637992 d
0.881408805031 3.46485433821 d
0.880754716981 3.49631751123 d
0.879628930818 3.56749003474 d
0.879006289308 3.60033219976 d
0.878694968553 3.61770219998 d
0.87770754717 3.67541596939 d
0.877650943396 3.67975112644 d
0.87806918239 3.65391690663 d
0.876849056604 3.72685085583 d
0.877468553459 3.6923780902 d
0.877163522013 3.71036646414 d
0.877194968553 3.70470853771 d
0.876924528302 3.7226677664 d
0.0279090909091 98.6352707315 e
0.0344090909091 98.6316243693 e
0.0386818181818 98.52492747 e
0.0475454545455 98.1874422417 e
0.0537727272727 97.8990921292 e
0.061 97.6861659579 e
0.0673181818182 97.3124890159 e
0.0761363636364 96.9698026742 e
0.0820454545455 96.7494134392 e
0.0961363636364 95.9512587918 e
0.103363636364 95.5940436858 e
0.111954545455 94.9767543403 e
0.113545454545 95.1375054002 e
0.127863636364 94.134937515 e
0.134863636364 93.6545693185 e
0.141318181818 93.1291825101 e
0.154727272727 92.1528613758 e
0.153545454545 92.2462304628 e
0.161636363636 91.7678360211 e
0.164363636364 91.6274627142 e
0.470782608696 53.2346828617 f
0.618739130435 33.441431878 f
0.699695652174 22.1324970265 f
0.746554347826 15.7984812585 f
0.759152173913 14.2687511469 f
0.775043478261 12.2558463446 f
0.781380434783 11.5354177469 f
0.787684782609 10.8703601254 f
0.793695652174 10.2447635926 f
0.795239130435 10.0963952516 f
0.796945652174 9.92881291058 f
0.800184782609 9.62768252272 f
0.801630434783 9.47720645348 f
0.80322826087 9.34143916869 f
0.804054347826 9.26564766576 f
0.806793478261 9.01033119491 f
0.807739130435 8.92790431267 f
0.808782608696 8.83243771404 f
0.809086956522 8.80154688328 f
0.809652173913 8.75267865253 f
0.210675675676 81.1080453955 g
0.268243243243 76.7647914962 g
0.325027027027 71.2645805719 g
0.363621621622 67.3281186014 g
0.393486486486 63.7827089521 g
0.423891891892 60.249370337 g
0.450351351351 56.9480749082 g
0.466648648649 54.7152534477 g
0.479432432432 52.8767395278 g
0.49372972973 50.8758254844 g
0.500594594595 49.8631253694 g
0.506594594595 48.9310219605 g
0.507648648649 48.7992837605 g
0.512324324324 48.0758996725 g
0.515567567568 47.5864283437 g
0.519243243243 47.0141374136 g
0.521027027027 46.7375687319 g
0.522918918919 46.4460466384 g
0.525972972973 45.9721508351 g
0.526945945946 45.8186923719 g
0.975576875769 0.691732673734 h
0.988682656827 0.120174981477 h
0.994094710947 0.0237731167541 h
0.99655104551 0.00801427789458 h
0.997009840098 0.00409815487216 h
0.997280442804 0.00315156707807 h
0.997467404674 0.00233863646107 h
0.997473554736 0.00225018331187 h
0.99767896679 0.00180486541674 h
0.997699876999 0.00170744031856 h
0.997749077491 0.0015738107167 h
0.997831488315 0.00143756263537 h
0.997833948339 0.00136290789115 h
0.997899138991 0.00128265135989 h
0.99786100861 0.00132557996441 h
0.997928659287 0.0012236731525 h
0.997923739237 0.00123263071092 h
0.997961869619 0.0011934361387 h
0.997960639606 0.00119455597215 h
0.99795202952 0.00120314189149 h
}\datatable

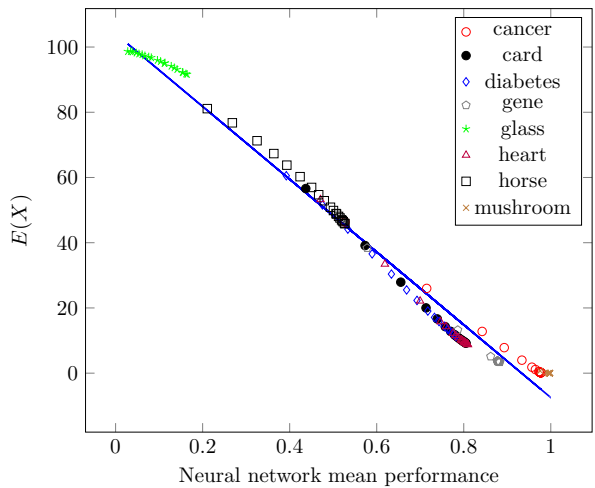
\begin{figure}[b]
 \center
 \begin{tikzpicture}[scale=0.8]
\begin{axis}[%
ylabel = $E(X)$,
xlabel = Neural network mean performance,
legend entries= {cancer, card, diabetes, gene, glass, heart, horse, mushroom},
scatter/classes={%
    a={mark=o,draw=red},
    b={mark=*,draw=black},
    c={mark=diamond,draw=blue},
    d={mark=pentagon,draw=gray},
    e={mark=star,draw=green},
    f={mark=triangle,draw=purple},
    g={mark=square,draw=black},
    h={mark=x,draw=brown}
	}]
\addplot[scatter,only marks,%
    scatter src=explicit symbolic]%
table[meta=label] {\datatable };
\addplot[thick, blue]table[y={create col/linear regression={y=Y}}]{\datatable};
\end{axis}
\end{tikzpicture}
\caption{Cancer, card, diabetes, gene, glass, heart, horse and mushroom datasets mean performance versus $E(X)$\label{fig:results}}
\end{figure}

Let $X$ be a random variable representing the number of 1s in the output of the probabilistic quantum memory.
Table~\ref{tab:resultcancer} shows the results of the experiment for cancer and gene datasets. We can verify that the expected value $E(X)$ is related to	 neural network mean performance. An increase in performance corresponds to a reduction in $E(X)$.

 In Fig.~\ref{fig:results} we plot the mean performance of each architecture versus the expected value of the $X$ for cancer, card, diabetes, gene, glass, heart, horse and mushroom datasets.
 We can easily see that there is an approximately linear relation between the neural network mean performance and $E(X)$.
 In this way, we can use Algorithm~1 to select a near-optimal neural network and also to estimate the mean-performance of the neural network architecture over a dataset.

 \begin{table}
\subfloat[Cancer dataset]{
\begin{tabular}{lll}
Neurons  & Performance & $E(X)$\\ \hline
1 & 0.5269 & 46.3695 \\ \hline
2 & 0.7148 & 25.9858 \\ \hline
3 & 0.8423 & 12.7966 \\ \hline
4 & 0.8930 & 7.8277 \\ \hline
5 & 0.9336 & 3.9903 \\ \hline
6 & 0.9566 & 1.8340 \\ \hline
7 & 0.9647 & 1.1213 \\ \hline
8 & 0.9730 & 0.4200 \\ \hline
9 & 0.9733 & 0.4945 \\ \hline
10 & 0.9753 & 0.3059 \\ \hline
11 & 0.9763 & 0.1581 \\ \hline
12 & 0.9764 & 0.1571 \\ \hline
13 & 0.9766 & 0.1543 \\ \hline
14 & 0.9773 & 0.1461 \\ \hline
15 & 0.9765 & 0.1543 \\ \hline
16 & 0.9769 & 0.1510 \\ \hline
17 & 0.9770 & 0.1488 \\ \hline
18 & 0.9774 & 0.1426 \\ \hline
19 & 0.9775 & 0.1423 \\ \hline
20 & 0.9772 & 0.1449 \\ \hline
\end{tabular}
}
\quad
\subfloat[Gene dataset]{
\begin{tabular}{lll}
Neurons  & Performance & $E(X)$\\ \hline
1 & 0.5778 & 38.3819 \\ \hline
2 & 0.7862 & 13.2412 \\ \hline
3 & 0.8619 & 5.1163 \\ \hline
4 & 0.8781 & 3.7319 \\ \hline
5 & 0.8809 & 3.5152 \\ \hline
6 & 0.8829 & 3.3877 \\ \hline
7 & 0.8827 & 3.3940 \\ \hline
8 & 0.8814 & 3.4649 \\ \hline
9 & 0.8808 & 3.4963 \\ \hline
10 & 0.8796 & 3.5675 \\ \hline
11 & 0.8790 & 3.6003 \\ \hline
12 & 0.8787 & 3.6177 \\ \hline
13 & 0.8777 & 3.6754 \\ \hline
14 & 0.8777 & 3.6798 \\ \hline
15 & 0.8781 & 3.6539 \\ \hline
16 & 0.8768 & 3.7269 \\ \hline
17 & 0.8775 & 3.6924 \\ \hline
18 & 0.8772 & 3.7104 \\ \hline
19 & 0.8772 & 3.7047 \\ \hline
20 & 0.8769 & 3.7227 \\ \hline
\end{tabular}
}
\caption{Results cancer dataset (left) and gene datset (right)\label{tab:resultcancer}}
  \end{table}

One limitation of classical neural networks is the absence of an algorithm to determine the best neural network architecture for a given dataset.
Classical strategies to evaluate neural network architectures requires a costly process that can last from minutes to days\cite{benardos2007optimizing}.
How to select a neural network architecture is yet an open problem and the use of more complex neural networks with deep architectures increases the complexity to determine a neural network architecture with optimal performance.

In this work, we explore the principles of quantum computing to create a hybrid classical and quantum algorithm to perform classical neural network architecture selection.
If $C$ is the cost to train one neural network, given $n$ neural networks architectures the proposed method has cost $O(n \cdot C)$ and
determine a near-optimal neural network architecture in the given set of architectures.
A classical algorithm to evaluate $n$ neural networks architectures over all possible initial weights will have at least cost $O(n\cdot 2^{|W|} \cdot C)$,
where $W$ is the set of all possible weights with a given precision. The proposed method has an exponential speed up when compared to its classical version.

One limitation of the proposed method is its inability to evaluate neural networks architectures with very close performance.
This limitation occurs because of the use of Hamming distance.
Then the method should be used to select a set of near-optimal neural networks and then a classical experimentation could be performed to finish the search for the neural network with the best performance.

\begin{figure}
 \center
 \begin{tikzpicture}[scale=0.8]
\begin{axis}[%
xlabel=Number of 1s in the output,
ylabel=Probability,
scatter/classes={%
    a={mark=none,draw=blue, dashed},
        b={mark=none,draw=red, solid}}]
\addplot[color=black,scatter,mark = none%
    scatter src=explicit symbolic, dashed]%
table[meta=label]
{
x y label
0 0.12000335063505205 a
1 0.034844989979120555 a
2 0.012982516778834672 a
3 0.006507358583242946 a
4 0.004159923361565486 a
5 0.0030115552829131746 a
6 0.002236028573355277 a
7 0.0016558311830489633 a
8 0.0012591961215696061 a
9 0.001040268796485861 a
10 0.0009698370394521496 a
11 0.001005821274348644 a
12 0.0011073545669515355 a
13 0.0012445270241857329 a
14 0.0014053707255416712 a
15 0.0016022893211334922 a
16 0.0018782704271757195 a
17 0.0023111801652640385 a
18 0.003012777282706537 a
19 0.0041183457586279225 a
20 0.005763911596330046 a
21 0.008051475127687785 a
22 0.011008039318608128 a
23 0.01454950469380434 a
24 0.018462962037791648 a
25 0.022418335013582768 a
26 0.02601251275920407 a
27 0.028838471448105973 a
28 0.03056258788732035 a
29 0.030989413500921093 a
30 0.030096463475634813 a
31 0.028030845878851983 a
32 0.02507110288222677 a
33 0.0215670872271734 a
34 0.017874967886763404 a
35 0.01430287639293937 a
36 0.01107687624024674 a
37 0.008329656458175594 a
38 0.006108249721192589 a
39 0.004393683283816132 a
40 0.0031249718423039933 a
41 0.00222145213557772 a
42 0.0015999620374308903 a
43 0.0011857628291360774 a
44 0.0009178049381915461 a
45 0.0007498000422976636 a
46 0.0006487374432566647 a
47 0.0005922365761782249 a
48 0.0005657089579000287 a
49 0.0005598817377560094 a
50 0.0005688993806273097 a
51 0.0005889999712670479 a
52 0.0006176561371565763 a
53 0.0006530594630015474 a
54 0.0006938832212098737 a
55 0.0007393446932569128 a
56 0.0007896636160921812 a
57 0.0008470370566214124 a
58 0.0009171927766159381 a
59 0.0010114299627264338 a
60 0.0011488189498303286 a
61 0.0013579505178353641 a
62 0.001677375033882599 a
63 0.0021537607498562177 a
64 0.0028369560852589282 a
65 0.003771666943862327 a
66 0.004986371348009711 a
67 0.006481244393482018 a
68 0.008217923210900244 a
69 0.010114444265274315 a
70 0.012048214483581971 a
71 0.013868279999116509 a
72 0.015415719043842669 a
73 0.01654843641077597 a
74 0.017164910033079784 a
75 0.017221277622185418 a
76 0.016737730404441025 a
77 0.015792983971443634 a
78 0.014508625506540843 a
79 0.013027379764292768 a
80 0.011490219261493871 a
81 0.010016807159302095 a
82 0.008692469827987844 a
83 0.00756325814728033 a
84 0.006638953801543227 a
85 0.0059022161487400976 a
86 0.005320668381842678 a
87 0.0048581186224688115 a
88 0.004481924708088155 a
89 0.004165832241427435 a
90 0.0038903844000208383 a
91 0.0036440926065412775 a
92 0.0034262856657410927 a
93 0.003248495566366989 a
94 0.003131408566024442 a
95 0.003102989869192961 a
96 0.003210758220403917 a
97 0.0035520759131945977 a
98 0.0043400456906077176 a
99 0.006388405070026672 a
100 0.10839522086465346 a
    };\addlegendentry{1}

    \addplot[color=red,scatter,mark = none,
    scatter src=explicit symbolic]%
table
{
x y label
0 0.8713988925267684 b
1 0.11602976465400285 b
2 0.011524472929307295 b
3 0.0009683556482894353 b
4 7.306909949931523e-05 b
5 5.09893318341769e-06 b
6 3.263699807469925e-07 b
7 1.882360252777071e-08 b
8 9.689546801096891e-10 b
9 4.449354430170217e-11 b
10 1.830765528682242e-12 b
11 6.792903446291629e-14 b
12 2.287744587034023e-15 b
13 7.036293063428191e-17 b
14 1.9872152090252653e-18 b
15 5.178594317883872e-20 b
16 1.2505263076471686e-21 b
17 2.8087626966820385e-23 b
18 5.887309383070745e-25 b
19 1.1549821522401019e-26 b
20 2.126323406368821e-28 b
21 3.682132543604013e-30 b
22 6.010400372642044e-32 b
23 9.26552046008985e-34 b
24 1.3512913172038848e-35 b
25 1.867335131297889e-37 b
26 2.448555137341666e-39 b
27 3.0505458103914597e-41 b
28 3.6152812509978134e-43 b
29 4.080151572800156e-45 b
30 4.389480263559621e-47 b
31 4.505563933540137e-49 b
32 4.4161923284451415e-51 b
33 4.136591752475239e-53 b
34 3.705426962317446e-55 b
35 3.1762440055734096e-57 b
36 2.6069003131599734e-59 b
37 2.049757027686957e-61 b
38 1.544752998374418e-63 b
39 1.116312632232666e-65 b
40 7.738475160041134e-68 b
41 5.147809849190042e-70 b
42 3.287190912583989e-72 b
43 2.01550637227829e-74 b
44 1.1868779586387532e-76 b
45 6.713999241555291e-79 b
46 3.6491007617316583e-81 b
47 1.9058181886042157e-83 b
48 9.565679145843746e-86 b
49 4.6144801604580956e-88 b
50 2.1395511068468216e-90 b
51 9.535033055765696e-93 b
52 4.084271927125916e-95 b
53 1.6814336207793842e-97 b
54 6.652479349012907e-100 b
55 2.5291721281933973e-102 b
56 9.238523840945856e-105 b
57 3.2417544422061536e-107 b
58 1.0924970843096624e-109 b
59 3.5352267543285975e-112 b
60 1.0981194693799361e-114 b
61 3.273252119724623e-117 b
62 9.359487858778413e-120 b
63 2.566225654844998e-122 b
64 6.743980445634577e-125 b
65 1.6978728213402592e-127 b
66 4.092879656082977e-130 b
67 9.441321777691349e-133 b
68 2.082751886664422e-135 b
69 4.390744177243995e-138 b
70 8.838966701640824e-141 b
71 1.6977123727213566e-143 b
72 3.1083457825374224e-146 b
73 5.419562216014152e-149 b
74 8.988680848478218e-152 b
75 1.4164712208916235e-154 b
76 2.1180386531738778e-157 b
77 3.0009178581859115e-160 b
78 4.022416627341969e-163 b
79 5.0919291327766827e-166 b
80 6.075910506033594e-169 b
81 6.819555294633015e-172 b
82 7.182828771698117e-175 b
83 7.080919160193097e-178 b
84 6.514168400631112e-181 b
85 5.573907386347594e-184 b
86 4.419287604847312e-187 b
87 3.232665431532706e-190 b
88 2.1708066794793883e-193 b
89 1.3304914822183534e-196 b
90 7.392000442391286e-200 b
91 3.692498281105708e-203 b
92 1.6420058265166136e-206 b
93 6.420686158238076e-210 b
94 2.173458431795753e-213 b
95 6.239915284750868e-217 b
96 1.477328835697281e-220 b
97 2.76926898545858e-224 b
98 3.8535414145935467e-228 b
99 3.538787966286079e-232 b
100 1.6086229176980981e-236 b
    }; \addlegendentry{19}
\end{axis}
\end{tikzpicture}
\caption{Probability output for cancer dataset with 1 hidden neuron and 19 hidden neurons \label{fig:219}}
\end{figure}
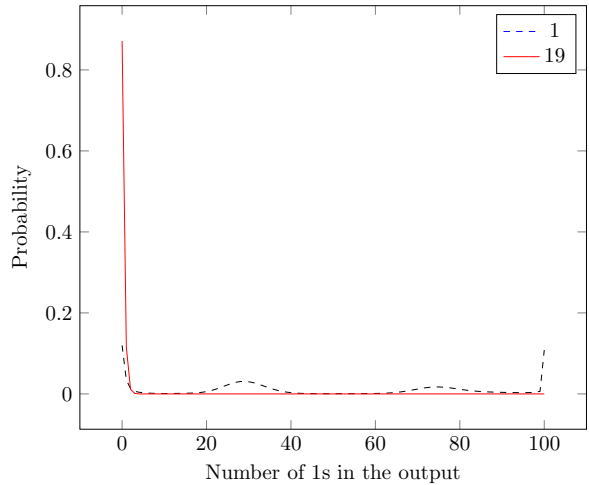

With a promise that the neural networks to be evaluated have a significant difference in their accuracy over the test set, the proposed method will perform an optimal selection (instead of near-optimal) with high probability. For instance, with 1 neuron in the hidden layer we obtain a neural network mean accuracy of 0.52 and with 19 neurons in the hidden layer, the neural network obtains mean accuracy of 0.97.  With the 19 hidden neurons neural network the proposed method will have 0 or 1 ones in the output with probability 0.9874 and the neural network with 1 hidden neuron will have 0 or 1 ones in the output with probability 0.1548. With the objective to illustrate the behavior  of the proposed method, Fig.~\ref{fig:219} presents the number of ones in the output of the  probabilistic memory and the probability of each output.

The neural network architecture selection uses quantum registers inputs, weights, desired output, calculated output and performance. Algorithm~1 creates a superposition of neural networks with all possible weights and evaluate the performance of each neural network in superposition. This evaluation in a quantum superposition is based on a neural network quantum learning algorithm\cite{panella2011neural}, one of the authors generalizes this strategy to perform a selection\cite{da2016quantum}\footnote{Where we supposed the viability of nonlinear quantum operators} and evaluation\cite{silva2017quantum} of neural networks architectures. We notice a related work where a quantum superposition of classifiers is used to perform parameter selection\cite{dunjkoadvances}. 

The main idea of the quantum cross-validation is to explore the quantum parallelism to execute a $\kappa$-fold cross-validation, training the model only once. This strategy leads to  a constant speedup in the cross-validation process. The exponential speedup obtained in this paper came from the superposition of neural networks. In this way, using cross-validation we can evaluate an exponential number of neural networks with the cost to train and run a single neural network.

\section{Conclusion}
\label{sec:conclusion}
In this work, we proposed a classical-quantum algorithm to select neural networks architectures (number of neurons in the hidden layer).  We evaluated the proposed method using its classical description and reducing the number of artificial neural networks in superposition.

Our main result is the ability to evaluate one neural network architecture through a $\kappa$-fold cross-validation with the cost to train only one neural network instance. The proposed method evaluates an exponential number of neural networks weights simultaneously with an exponential improvement in computational cost when compared with known classical alternatives.
The fast neural network evaluation allows the selection of near-optimal neural network architectures by repeating the cross-validation for each neural network architecture.

Quantum computation can be used to evaluate neural networks models  with an exponential speedup. 
The lack of experimentation is one problem in quantum machine learning because of the non-existence of quantum computers with enough quantum bits. 
To allow experimentation, we use benchmark problems to evaluate (without loss of generalization) a classical simplified  version of the proposed method.

One possible future work is to extend the proposed method to deal with models with close performance. 
One suggestion to accomplish this improvement is to change the probabilistic associative memory to deal with others distance functions. 
We also can use other kinds of quantum memories and machine learning models.

\section*{Acknowledgement }
This work was supported by the Serrapilheira Institute (grant number Serra-1709-22626), CNPq and FACEPE (Brazilian research agencies).

\bibliographystyle{unsrt}
\bibliography{bibliografia}

\end{document}